\documentclass[twocolumn,showpacs,amsmath,amssymb,prb,superscriptaddress]{revtex4}

\usepackage{graphicx}
\usepackage{bm}
\begin{document}

\title{Topological Fermi arcs in superfluid $^3$He}

\author{ M.A. Silaev } \affiliation{ Institute for Physics of Microstructures RAS, 603950 Nizhny Novgorod, Russia}
\author{ G.E. Volovik } \affiliation{ Low
Temperature Laboratory, Aalto University, P.O. Box 15100, 00076
Aalto, Finland} \affiliation{ L.~D.~Landau Institute for
Theoretical Physics, 117940 Moscow, Russia}

\date{\today}

\begin{abstract}
We consider fermionic states bound on domain walls in a Weyl
superfluid $^3$He-A and on interfaces between $^3$He-A and a fully
gapped topological superfluid $^3$He-B. We demonstrate that in
both cases fermionic spectrum contains Fermi arcs which are
continuous nodal lines of energy spectrum terminating at the
projections of two Weyl points to the plane of surface states in
momentum space. The number of Fermi arcs is determined by the
index theorem which relates bulk values of topological invariant
to the number of zero energy surface states. The index theorem is
consistent with an exact spectrum of Bogolubov- de Gennes equation
obtained numerically meanwhile the quasiclassical approximation
fails to reproduce the correct number of zero modes. Thus we
demonstrate that topology describes the properties of exact
spectrum beyond quasiclassical approximation.
\end{abstract}

\maketitle

\section{Introduction}

Chiral Weyl fermions represent the fermionic sector in the
Standard Model of particle physics. The point nodes in the
spectrum of chiral quarks and leptons are topologically protected,
as well as their condensed matter counterparts, which are called
Dirac or Weyl points (on topology of Weyl points in particle
physics and condensed matter see e.g.
 Refs. \onlinecite{Froggatt1991,Volovik2003,Horava2005,Kaplan2011,Zubkov2012b}).

The nodal topological systems with Weyl fermions  demonstrate different types of the bulk-surface and the bulk-vortex correspondence.
  Due to the bulk-vortex correspondence, the cores of some vortices in the Weyl superfluid $^3$He-A
    contain a dispersionless branch of bound states with zero
  energy which is one-dimensional flat band, discussed first by Kopnin and Salomaa in 1991 \cite{KopninSalomaa1991}.
  The end points of this flat band are determined by the projections of Weyl points to the direction of  vortex axis
(see Refs. \onlinecite{Volovik2010,MengBalents2012} for the topological origin of this flat band, and Refs.
\onlinecite{HeikkilaKopninVolovik2011,Ryu2002,SchnyderRyu2010,VolovikQuantumVacuum}
 for discussion of the topological flat bands in general). Due to the bulk-surface correspondence,
  the surface of a system with Weyl points  contains another exotic
  object -- the Fermi arc -- the 1D Fermi line in the 2D momentum space, which terminates on
  the projections of  two Weyl points to the plane of the surface.
  The Fermi arc on the surface of $^3$He-A shown in Fig.\ref{Fig:BoundaryStates}a has been considered in Ref. \onlinecite{Tsutsumi2011}
and that on the surface of topological semi-metals with Weyl
points -- in Refs.
\onlinecite{XiangangWan2011,Burkov2011,Hosur2012}.
  The flat band in the vortex core and the Fermi arc on the surface are momentum-space analogs of the
  real-space Dirac string terminating on two magnetic monopoles \cite{HeikkilaKopninVolovik2011}.

Fermi arcs appear also at the interface, which separates two
degenerate states  of $^3$He-A with the opposite directions of the
orbital anisotropy vector ${\bf \hat l}$ (in chiral superfluid
$^3$He-A vector ${\bf \hat l}$  determines the orientation of a
spontaneous  orbital angular momentum  of this anisotropic liquid
and also determines the positions ${\bf p}=\pm p_F{\bf \hat l}$ of
two Weyl points on the Fermi surface with topological charges
$N_3=\pm 2$, if spin degeneracy is taken into account). Bound
states emerging at one of the representatives of such interface --
the continuous $\hat{\bf l}$-soliton -- have been calculated in
Ref. \onlinecite{Wilczek}. Here we consider bound states emerging
on the singular domain wall in $^3$He-A discussed in Refs.
\onlinecite{Ohmi1982,SalomaaVolovik1989}. Zero energy edge states
on such domain wall in a thin film of $^3$He-A and the topological
bulk-edge correspondence for this 2+1 system were considered in
Ref. \onlinecite{Volovik1992}. Fermi arcs emerging in the  3+1
system  form a special configuration in momentum space, see Fig.
\ref{Fig:BoundaryStates}b. We also consider Fermi arcs emerging at
the interface between two topologically different quantum vacua:
the $^3$He-A with Weyl points and the fully gapped $^3$He-B, which
also has nontrivial topology in momentum space
\cite{SalomaaVolovik1988,Schnyder2008,Volovik2009b,VolovikQuantumVacuum}
(AB interface).

 The Fermi arcs in superfluid $^3$He solitons studied in the present paper
 which differ qualitatively from that existing in other
systems such as the surface of $^3$He-A and Weyl semimetals. The
reason is that they can be obtained ultimately beyond the
quasiclassical approximation. It was found that the quasiclassical
approximation yields a large number of spectrum branches which
intersect the Fermi level.\cite{Wilczek,Nakahara} The number of
such branches depends on the parameters of the solition in
contradiction with the topological index theorem. We show that the
reason for this contradiction is that the quasiclassical
Bogoluibov - de Gennes (BdG) equations inherently miss the normal
reflection of quasiparticles from the spatially inhomogeneous
superfluid order parameter and take into account only Andreev
reflection. Thus the system of quasiclassical BdG equations is
often called the system of Andreev equations. As we will discuss
below the normal reflection is crucial to describe the Fermi arcs
in superfluid $^3$He solitons. In this case fermionic zero energy
states are realized on the quasiparticle trajectories passing at
the small sliding angle to the soliton plane. For such
trajectories the normal reflection is strongly enhanced and change
qualitatively the behavior of spectral branches near the zero
energy points on Fermi arcs. Here we find numerically exact
solutions of the spectral problems for BdG equations which
confirms the topological index theorem predictions. Thus our
results demonstrate that topological bulk-edge correspondence
describe the properties of the exact
 spectrum of localized boundary states which in some cases are
 missed in the quasiclassical approximation.

 The structure of the paper is as follows.
 In section \ref{Sec:Model} we introduce the model which is the
 BdG equations and the order parameter structures of domain wall
 in A phase
 and AB interface. The results are presented in section \ref{Sec:Results}
 where we consider at first topological nature of Fermi arcs and
 then the calculations of fermionic spectrum. We discuss the
 failure of quasiclassical approximation to describe Fermi
 arcs in the fermionic spectrum of domain wall  in A phase and AB interface
 in section \ref{Sec:Discussion}. The conclusion is given in section
 \ref{Sec:Conclusion}.

  \section{The model}
 \label{Sec:Model}

We calculate the spectrum of eigenstates of Bogolubov-de Gennes
(BdG) Hamiltonian describing spin-triplet $p$-wave
superfluids/superconductors
\begin{equation}\label{Eq:Hamiltonian}
\hat H=[\varepsilon(\hat p)-\mu]\hat\tau_3+\hat\tau_1 {\rm Re}
\hat\Delta- \hat\tau_2 {\rm Im} \hat\Delta,
\end{equation}
where $\varepsilon(\hat p)=(\hat p_x^2+\hat p_y^2+\hat p_z^2)/2m$,
$\mu$ is chemical potential, $\hat p_i= -i \nabla_i$ and
 $\hat\tau_i$ are Pauli matrices of Bogolubov--Nambu spin
 acting on the wave function $\psi=(u,v)^T$ with particle and hole components $u$ and $v$.
 The gap operator is $\hat\Delta=A_{\alpha i}\hat\sigma_\alpha \hat p_i/p_F$, where $\hat\sigma_\alpha$ are
Pauli matrices of  spin (in $^3$He it is nuclear  spin),
$p_F=\sqrt{2m\mu}$ is the Fermi momentum.

The order parameter in superfluid $^3$He is $3\times3$ matrix
$A_{\alpha i}$ where the Greek and Latin indices correspond to the
spin and orbital variables. We will consider the fermionic modes
localized on different 1D order parameter solitons, namely the
domain wall of a $^3$He-A and the AB interface with order
parameter inhomogeneity along the vector ${\bf \hat{n}}={\bf
\hat{x}}$ so that the momentum projections $p_y$ and $p_z$ are
conserved. At first we introduce the order parameter structure
corresponding to the discussed solitons.

 \begin{figure}[h!]
 \centerline{\includegraphics[width=1.0\linewidth]{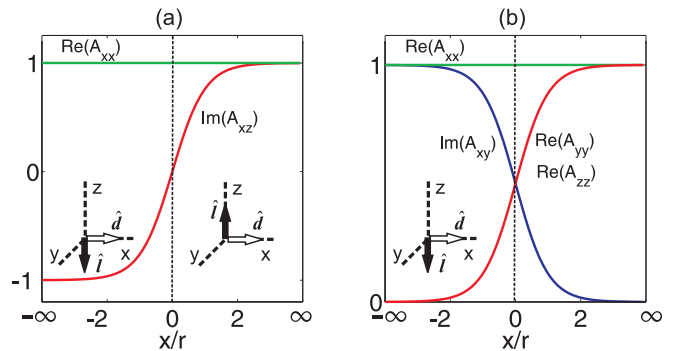}}
 \caption{\label{Fig:DWplots} (Color online)
 Model spatial dependencies of the order parameter components
 corresponding to the most symmetric and energetically preferable
 1D topological defects in superfluids $^3$He according to
 Refs.(\onlinecite{Ohmi1982,SalomaaVolovik1989,Salomaa1988,Schopohl}):
 (a) domain wall of the A phase and (b) AB
 interface. The order parameter is normalized to the bulk value. }
 \end{figure}

{\bf Domain wall of A phase.}
 We consider the configuration of one of the domain walls  possible in $^3$He-A according to the symmetry
 classification\cite{SalomaaVolovik1989}-- the one which is shown in Fig.
 \ref{Fig:DWplots}a.
 The orbital vector ${\bf \hat{l}}$ of the order parameter points down (${\bf \hat{l}} \parallel - {\bf
 \hat{z}}$) as $x\rightarrow -\infty$ and points up (${\bf \hat{l}} \parallel  {\bf
 \hat{z}}$) as $x\rightarrow + \infty$. This domain wall configuration can be
 approximated by the following Ansatz:
 \begin{equation}\label{Eq:AphaseSol}
  A_{\alpha i} (x) =\Delta_0 \hat d_\alpha [\hat x_i+i f(x) \hat y_i] \,,
 \end{equation}
 where the unit vector  $\hat {\bf d}$
 represents the spin part of the order parameter, which is fixed if the spin-orbit interaction is neglected,
 and we put $\hat {\bf d}=\hat {\bf x}$.
 Then the gap function is given by
 \begin{equation}\label{Eq:GapA}
 \hat\Delta_{AA}= c_0\hat\sigma_x\left[\hat p_x + i f(x)  p_y
 \right] \,,
 \end{equation}
 where $c_0=\Delta_0/p_F$ is the parameter which plays the role of
 the longitudinal speed of light in bulk $^3$He-A
 \cite{Volovik2003}, $f(x)$ is arbitrary monotonic function with asymptotic
 $f(\pm \infty)= \pm 1$.
 The topology of bound states does not depend
 on the details of the function $f(x)$ in Eq.(\ref{Eq:GapA}) if it is monotonic.
  Note that the same model (modulo unimportant ${\bf \hat d}$ vector orientation) describes domain wall between chiral
 domains in $p+ip$ superconductors like $Sr_2RuO_4$\cite{SigristDW} thus the
 fermionic spectra are identical in these cases.

{\bf AB interface.} Let us fix the order parameter on the B phase
side of the interface, i.e. at $x=+\infty$, in the following form
  \begin{equation}\label{Eq:Bphase}
 A_{\alpha i} (x=+\infty) =\Delta_B ( \hat x_\alpha \hat x_i + \hat y_\alpha \hat y_i + \hat z_\alpha \hat z_i).
 \end{equation}
 The configuration of the $A$ phase at $x\rightarrow - \infty$ can be
 different for different realizations of the domain wall (see e.g. Ref. \onlinecite{Salomaa1988}).
 They are described by the relative orientations of the vectors ${\bf \hat{l}}$, ${\bf \hat{d}}$
 and ${\bf \hat{n}}$ (the normal to the AB interface which in our case is ${\bf \hat{n}}={\bf \hat{x}}$).
First we shall consider the case ${\bf \hat{d}}= {\bf \hat{x}}$,
${\bf \hat{l}}= {\bf \hat{z}}$ which has the
 lowest energy \cite{Schopohl}
 \begin{equation}\label{Eq:Aphase}
 A_{\alpha j} (x=-\infty) =\Delta_A \hat x_\alpha (\hat x_j-i \hat y_j).
 \end{equation}
 We model the domain wall at the AB interface with $\Delta_A=\Delta_B=\Delta_0$ and
  the switching between bulk phases (\ref{Eq:Bphase},\ref{Eq:Aphase}) as follows
 \begin{eqnarray}\label{Eq:ABinterface}
 A_{xx}=\Delta_0 = const \\
 A_{yy}=A_{zz}=\Delta_0f_1(x) \\
 A_{xy}=-i\Delta_0 f_2(x)
 \end{eqnarray}
 where $f_{1,2}(x)$ are arbitrary monotonic functions with
 asymptotics
 $f_{1}(- \infty)= 0$, $f_{1}(+ \infty)= 1$ and $f_{2}(- \infty)= 1$, $f_{2}(+ \infty)=
 0$. The order parameter components (\ref{Eq:ABinterface})
 are shown in Fig.(\ref{Fig:DWplots}b).
  Then the gap operator is given by
 \begin{equation}\label{Eq:GapAB}
 \hat\Delta_{AB}= c_0\hat\sigma_x\left[\hat p_x - i f_2(x)  p_y\right]+
 c_0f_1(x)\left[\hat\sigma_y p_y +
 \hat\sigma_z p_z\right].
 \end{equation}

\section{Results}
\label{Sec:Results}

\subsection{Topology of Fermi arcs in superfluid $^3$He} \label{Subsec:Topology}

 \begin{figure}[h!]
 \centerline{\includegraphics[width=1.0\linewidth]{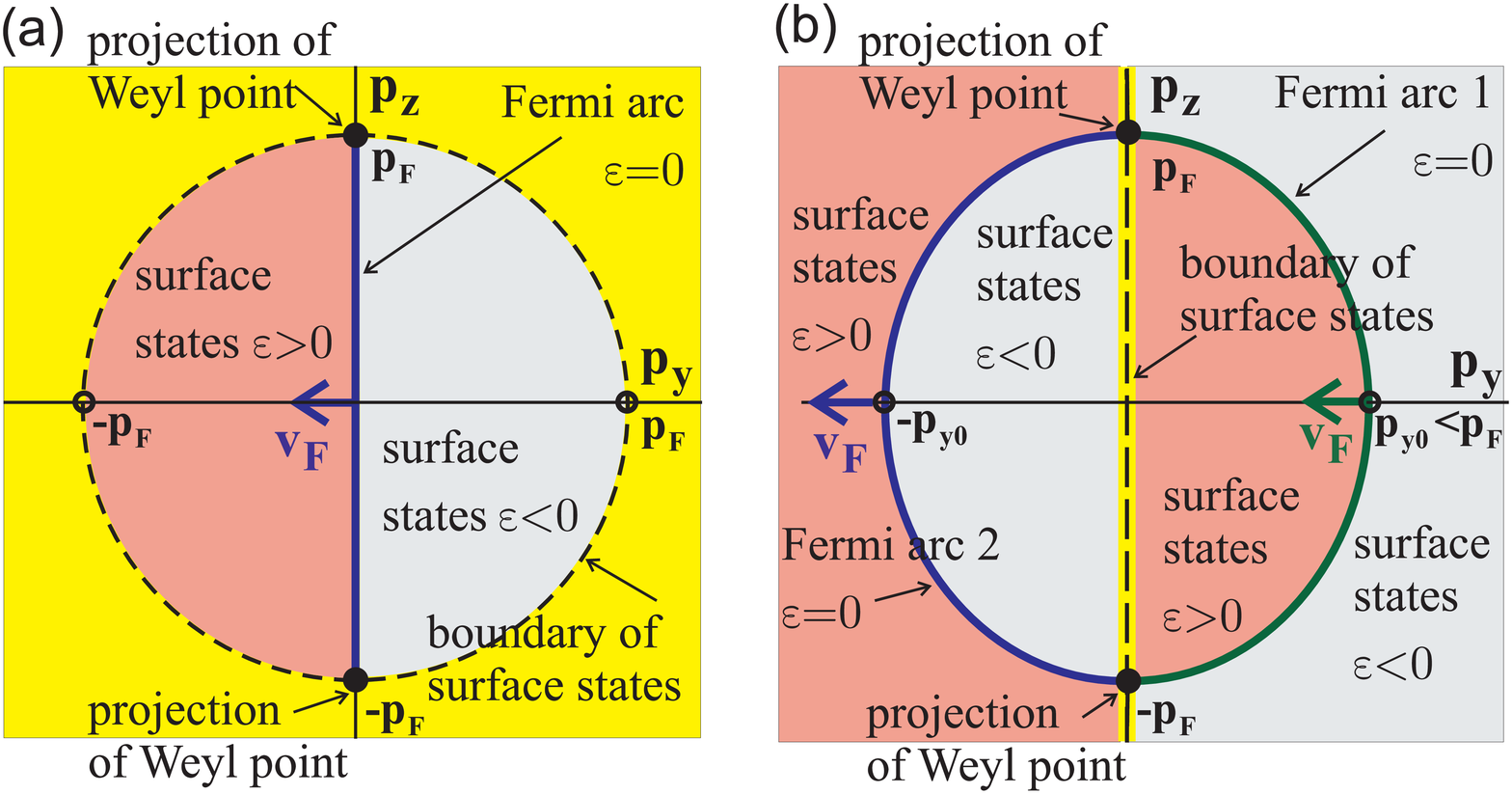}}
 \caption{\label{Fig:BoundaryStates} (Color online)
 The manifold of zero energy states in $p_y,p_z$ plane in
 the spectrum of bound fermions $\varepsilon(p_y,p_z)=0$ forms (a) the Fermi arc
 (solid blue line) on the boundary of $^3$He-A according to Ref.
 \onlinecite{Tsutsumi2011};
 (b) two Fermi arcs on the domain wall in Eq.(\ref{Eq:AphaseSol}).
 Only single spin projection is considered.
 Thick arrows show directions of Fermi velocity at the Fermi
 arcs. In (a) the Fermi arc has topological
 charge $N=+1$, which satisfies the index theorem following from the
 bulk-surface correspondence and the
 momentum space topology of Weyl points in bulk $^3$He-A. Fermi arc terminates on the projections
 of Weyl points to the surface of $^3$He-A.
 The spectrum of bound states terminates at the dashed line where the spectrum merges with the bulk spectrum.
 The region of continuous spectrum is shown by yellow shading.
 In (b) at $p_z=0$ the Fermi velocity at the Fermi
 arcs is in the same direction, ${\bf v_F}=-v_F {\bf y}$
 which demonstrates that both Fermi arcs have the same topological
 charge $N=+1$, which together satisfy the index theorem
 $\tilde{N}_3(left)-\tilde{N}_3(right)=+2$, in agreement with
 momentum space topology of Weyl points in bulk $^3$He-A on two
 sides of the domain wall in Fig. \ref{Arc}. Fermi arcs terminate on the projections
 of Weyl points to the interface.  The spectrum of bound states has discontinuity at
 $p_y=0$ where the spectrum merges with the bulk spectrum (the edge of continuum is shown by dashed line).  }
 \end{figure}

\begin{figure}[h!]
\centerline{\includegraphics[width=1.2\linewidth]{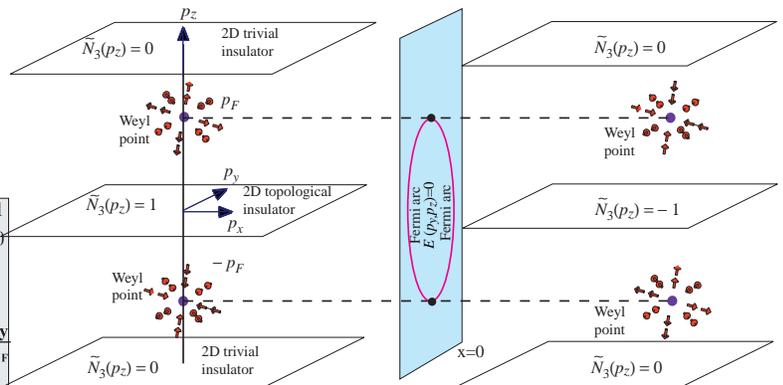}}
\caption{Topology of the fermionic bound states on the domain wall
in $^3$He-A. Momentum space topology of Weyl points in bulk
$^3$He-A on two sides of the wall prescribes existence of the
Fermi arcs in the spectrum of the fermionic states in the soliton
or at the interface between the bulk states with different
positions of Weyl points. In the considered case of the domain
wall the Weyl points on two sides of the interface have the same
positions in momentum space, but opposite topological invariants
$N_3$. This leads to two Fermi arcs terminating on the projections
of the Weyl points to the plane of domain wall (Fig.
\ref{Fig:BoundaryStates}b) according to the index theorem $\tilde
N_3({\rm left}) -\tilde N_3({\rm right})=2$. This is distinct from
the single Fermi arc on the surface of the $^3$He-A in Fig.
\ref{Fig:BoundaryStates}a. } \label{Arc}
\end{figure}
 As we have mentioned in the Introduction the
 spectrum of fermions bound to the surface of superfluid $^3$He
 A is known to contain the Fermi arc \cite{Tsutsumi2011}.
Let us now compare the Fermi arcs on a domain wall in $^3$He-A
 with the Fermi arc on a surface of $^3$He-A shown in Figs.(\ref{Fig:BoundaryStates}a) and
(\ref{Fig:BoundaryStates}b) correspondingly.
 In both cases Fermi arcs originate from the Weyl points in the bulk spectrum. But in case of the domain
 wall the Weyl points exist in the bulk liquids on both sides
 of the wall, as a result there are two Fermi arcs in Fig. \ref{Fig:BoundaryStates}b instead of a single Fermi arc in Fig.
 \ref{Fig:BoundaryStates}a. Topological origin of two Fermi arcs on the domain wall is demonstrated in Fig. \ref{Arc}.
 The Weyl points on two sides of the interface have the same
 positions in momentum space, but opposite values of topological
 invariant $N_3$.

 The bulk spectrum in the plane with fixed $p_z$
 in momentum space has no nodes if $|p_z| \neq p_F$ and thus
 corresponds
 to the spectrum of 2D insulator. This insulator is topological for $|p_z| < p_F$,
 since it is described by nonzero topological invariant introduced
 in Refs.
 \onlinecite{So1985,IshikawaMatsuyama1986,IshikawaMatsuyama1987,VolovikYakovenko1989}
 for 2+1 systems:
 \begin{equation}
 \begin{split}
 & \tilde N_3(p_z)
 \\
 & =\frac{1}{4\pi^2}~ {\bf tr}\left[  \int    dp_xdp_yd\omega
 ~G\partial_{p_x} G^{-1} G\partial_{p_y} G^{-1}G\partial_{\omega}
 G^{-1}\right]\,.
 \end{split}
 \label{N2+1}
 \end{equation}
 Here $G$ is the Green's function matrix, which in our
non-interacting models is $G=\left(i\omega -\hat H\right)^{-1}$.
One has $\tilde N_3(|p_z| < p_F) =+1$ on one side of the wall and
$\tilde N_3(|p_z| < p_F) =-1$ on the other side. According to the
index theorem \cite{Volovik1992,Volovik2003}, the difference
between these two values determines the number of the zero modes
at the interface between the 2+1 topological insulators for each
$|p_z| < p_F$. As a result one has two Fermi arcs at the soliton
wall.

\subsection{Calculation of the spectrum}
\label{Subsec:SpectrumCalculation}

 At first we note that the spectral problem for the domain wall can be
significantly simplified since the BdG Hamiltonian
(\ref{Eq:Hamiltonian}) with the gap operator given by
Eq.(\ref{Eq:GapA}) is proportional to the fermionic spin
 $\hat\sigma_x$. Thus we transform the quasiparticle wave function $\psi = (u, v)^T$ to remove the spin
  dependence of the order parameter $\tilde{u}=\hat\sigma_x u$,
  $\tilde{v}=v$ which yields the gap operator in the form
 \begin{equation}\label{Eq:GapOpNew}
 \hat\Delta= c_0\left[\hat p_x + i f(x) p_y\right].
 \end{equation}

 Furthermore the spectral problem for the AB interface with the gap operator given by
 Eq.(\ref{Eq:GapAB}) can be mapped on the case of
domain wall of the A phase. Let us transform the quasiparticle
wave function  components as follows $\tilde{u}=\hat\sigma_x u$,
$\tilde{v}=v$. We choose the spin basis $\chi_{\sigma} = (ip_z,
\sigma_\perp p_\perp-p_y)$
 of the eigenstates of the operator $\hat \sigma_\perp = \hat\sigma_zp_y-\hat\sigma_yp_z$
 corresponding to the eigenvalues $\sigma_\perp=\pm 1$ where $p_\perp=\sqrt{p_y^2+p_z^2}$. Then
 the order parameter is diagonal in spin space and has the form of Eq.(\ref{Eq:GapOpNew})
 where
 \begin{equation}\label{Eq:f}
 f(x)=\sigma_\perp p_\perp f_1 (x) - p_yf_2(x).
 \end{equation}
 In order to obtain the bound states the function (\ref{Eq:f}) has to satisfy the condition
 $f(+\infty)f(-\infty)<0 $ which yields $\sigma_\perp p_y>0$.
 Thus in contrast to the case of the domain wall considered above
  the spin degeneracy is removed since the proper spin state is
 determined by the condition of bound state existence.

 The continuous part of the spectrum $\varepsilon_c ({\bf p})$ of
 fermionic excitations is determined by the eigen states of
 Hamiltonian (\ref{Eq:Hamiltonian}) at the bulk regions $|x| \gg r$
 \begin{equation}\label{Eq:Edge}
 \varepsilon_c ({\bf p})=\pm
 \sqrt{[\varepsilon(p)-\mu]^2+c_0^2(p_x^2+p_y^2)}.
 \end{equation}
 The states localized at the domain wall are characterized by
 a single discrete quantum number which enumerates the energy branches
 and by the two continuous quantum numbers which are the
 projections of the quasiparticle momentum $p_{y,z}$ onto the domain wall plane. The energy
 of localized states is confined within the region
 $|\varepsilon|<\min_{p_x}\varepsilon_c ({\bf p})$.

 In general the eigenvalue problem of BdG Hamiltonian (\ref{Eq:Hamiltonian}) yields
 a system of differential equations which can not be solved
 analytically. However there are approximate methods which can
 help to study qualitative features of the spectrum.
 First of all we will employ the semiclassical approach\cite{RakhmanovAzbel}
 when the momentum operator is approximated by a number $\hat p_x=p_x(x)$
thus turning the differential BdG equations into the algebraic
ones. In this case the spectrum of bound states is determined by
Bohr-Sommerfeld quantization of classical periodic motion between
 reflection points discussed in the Section
\ref{Subsubsec:Semiclassical}. The semiclassical approximation
fails when the distance between reflection points is too small.
Particularly important is the case when two Andreev reflection
points come close together forming the bound state which
correspond to the so called 'zeroth branch' of the energy
spectrum\cite{Wilczek, Nakahara}. This case can be treated with
the help of another approach - the quasiclassical approximation
when one can use reduced order Andreev equations for the envelope
wave functions discussed in Section
\ref{Subsubsec:quasiclassical}. The quasiclassical approximation
in BdG equations is applicable as long as the normal reflection
can be neglected. For the structures considered in the present
paper the normal reflection is crucial to describe properly the
zero modes of the fermionic spectrum. This can be done only with
the help of numerical solution of the exact BdG system of
equations discussed in Section\ref{Subsubsec:ExactSp}.

Hereafter we will choose the model form of Eq.(\ref{Eq:GapOpNew})
with $f(x)=\tanh(x/r)$ where $r\sim \xi$ is width of
 the domain wall and $\xi=\hbar v_F/\Delta_0$ is the coherence length where $v_F=p_F/m$
 is the Fermi velocity. The fermionic spectrum in this configuration was
 considered in Ref. \onlinecite{Nakahara} in quasiclassical
 approximation. Here  in Sec. \ref{Subsubsec:ExactSp} we will implement an analysis of the fermionic spectrum
 beyond the quasiclassics in order to study the the Fermi arcs supporting Majorana states.

 \subsubsection{Semiclassical spectrum.}
 \label{Subsubsec:Semiclassical}
 We begin the analysis of the spectrum of localized states
 from the semiclassical approximation. Provided the condition $|p_x| \xi \gg 1 $ is valid
 we use the expression for the energy (\ref{Eq:Edge})
 substituting the local value of the order parameter. In this case we obtain
 \begin{equation}\label{Eq:px}
 \varepsilon =\pm
 \sqrt{[p_x^2/2m-\mu_x]^2+c_0^2[p_x^2+f^2(x) p_y^2]}.
 \end{equation}
 where
 $\mu_x=\mu-(p_y^2+p_z^2)/2m$.
 From Eq.(\ref{Eq:px}) the function $p_x=p_x(x)$ can be found which
 describes semiclassical orbits in $(p_x,x)$ phase space.
 In general the orbits can have two types of stationary points
 determined by the nature of quasiparticle reflection.

  The normal reflection occurs at the points $x=x^n_{1,2}$ defined by $p_x
 (x^n_{1,2})=0$ or equivalently
 \begin{equation}\label{Eq:NormalRef}
 c_0p_y f(x^n_{1,2})= \pm\sqrt{\varepsilon^2-\varepsilon_{n2}^2}
 \end{equation}
 and exists at the energy interval
 $\varepsilon_{n1}>|\varepsilon|>\varepsilon_{n2}$
 where
 $\varepsilon_{n1}=\sqrt{(c_0p_y)^2+\mu_x^2}$ and
 $\varepsilon_{n2}=|\mu_x|$.
 The Andreev reflection occurs at $x=x^a_{1,2}$ where $Im(p_x)$
 becomes non-zero
 \begin{equation}\label{Eq:AndreevRef}
 c_0p_yf(x^a_{1,2})= \pm  \sqrt
 {\varepsilon^2-\varepsilon_{a2}^2}
 \end{equation}
 and exists if
 $\varepsilon_{a1}>|\varepsilon|>\varepsilon_{a2}$
 where
 $\varepsilon_{a1}=c_0\sqrt{p_y^2+2m\mu_x-(mc_0)^2}$ and
 $\varepsilon_{a2}=c_0\sqrt{2m\mu_x-(mc_0)^2}$.

\begin{figure}[htb!]
 \centerline{\includegraphics[width=1.0\linewidth]{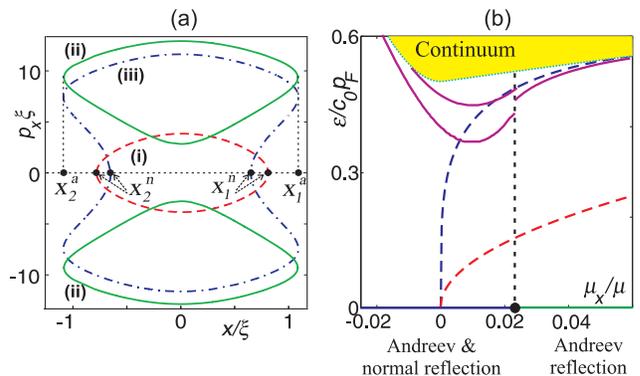}}
 \caption{\label{Fig:orbit} (Color online) (a) Closed orbits $p_x=p_x(x)$ determined by the semiclassical Eq.(\ref{Eq:px})
 at the domain wall for $p_z=0$, $p_y=0.5 p_F$ and $c_0 /v_F= 2.5\;10^{-3}$. Shown by
 red dashed, green solid and blue dash-dotted curves are
 the regimes ${(\bf i)}$, ${(\bf ii)}$ and ${(\bf iii)}$ discussed
 in the text. The positions of Andreev $x_{a1,2}$ and normal
 $x_{n1,2}$ reflection points on the semiclassical orbits is
 indicated.
 (b) Comparison of energy spectrum given by quasiclassical  and semiclassical
 approximations. The quasiclassical spectrum  given by Eq.(\ref{Eq:Nakahara}) is shown for $n_q=0,1$ by dashed red and blue
 lines. Quasiclassical approximation takes into account only Andreev reflection hence the spectrum is doubly degenerate by $sign(p_x)$.
 The semiclassical approach describes both Andreev and normal reflection
 hence the degeneracy by $sign(p_x)$ is removed.
 The branches given by Bohr-Sommerfeld quantization (\ref{Eq:BS}) of semiclassical orbits with $n_s=1,2$ are shown by solid magenta lines.
 The number of semiclassical branches is doubled due to the normal reflection
 when the Andreev orbits shown in the panel (a) by green solid lines merge into the single one shown by blue dash-dotted
 line with both normal and Andreev reflection points. }
 \end{figure}

 One can see that $\varepsilon_{n1}>\varepsilon_{a1}$
 and $\varepsilon_{n2}>\varepsilon_{a2}$ therefore Eq.(\ref{Eq:px}) determines
 two qualitatively different types of the enclosed classical orbits
 in $(p_x,x)$ space. That is for the energies ${\bf (i)}$ $\varepsilon_{n1}>|\varepsilon|>\varepsilon_{a1}$
 the orbits have only normal reflection points and
 for ${\bf (ii)}$ $\varepsilon_{n2}>|\varepsilon|>\varepsilon_{a2}$
 only Andreev reflection points. The orbits of type ${\bf (i)}$ and
 ${\bf (ii)}$ are shown in the Fig.(\ref{Fig:orbit}a) by red dashed and green solid lines correspondingly.
 Provided ${\bf (iii)}$ $\varepsilon_{a1}>\varepsilon_{n2}$ there is also the
 third regime when the orbit has both normal and Andreev
 reflection points which is shown by the blue dash-dotted line in the
 Fig.(\ref{Fig:orbit}).
 The energy spectrum is determined by the Bohr-Sommerfeld
 quantization
 \begin{equation}\label{Eq:BS}
 \oint p_x dx =2\pi (n_s+\gamma)
 \end{equation}
 where $n_s$ is integer, $\gamma= 1/2$ for the orbits with two normal reflection points and
  $\gamma=0$ for the orbits without normal reflection points \cite{RakhmanovAzbel}.
   Semiclassical spectral branches for
 $n_s=1,2$ are shown in Fig.(\ref{Fig:orbit}b) by solid lines.
 The number of branches is doubled when the classical orbits with only Andreev reflection points merge into
 the single one with both normal and Andeev reflection points. The discontinuity of semiclassical branches
 in Fig.\ref{Fig:orbit}(b) is caused by the emergence of two normal
 reflection points on the classical orbits which change abruptly the factor $\gamma$ in Bohr-Sommerfeld quantization rule.

 From the above semiclassical consideration we obtain that the energy of localized states is not bounded as function
 of $p_y$ and $p_z$. Indeed let us consider the case {\bf (i)} and
 assume $\sqrt{\varepsilon^2-\mu_x^2}\ll c_0|p_y|\ll|\varepsilon|$. In this
 case we can put $f(x)= x/r$
 in the Eq. (\ref{Eq:px}) which yields then the harmonic oscillator
 spectrum
\begin{equation}\label{Eq:HO}
\varepsilon_{ns} (p_y,p_z)=  |\mu_x| + \omega (n_s+1/2)
\end{equation}
 where $\omega=c_0(|p_y|/p_Fr)\sqrt{2\mu/|\mu_x|}$. One
 can see that the Eq.(\ref{Eq:HO}) yields $\varepsilon>  |\mu_x|$.
 Hence the energy branches at $\mu_x<0$ can have arbitrary high
 energy when $\mu_x\rightarrow -\infty$. This tendency can be seen in the Fig.(\ref{Fig:orbit}b).

Besides the energy branches determined by Bohr-Sommerfeld
quantization with $n_s\geq 1$ there exists a so-called 'zeroth
branch' of the spectrum which contains the zero modes
$\varepsilon=0$. This branch can not be obtained from the
semiclassical approach therefore the different treatment is
needed\cite{Wilczek}.

  \subsubsection{Quasiclassical approximation.}
  \label{Subsubsec:quasiclassical}

  To find the zeroth branch analytically
  we assume the restriction $(p_x/p_F)p_x\xi\gg 1$
 where $p_x=\sqrt{2m\mu_x}$ is a constant number.
 Qualitatively this restriction corresponds to the regime
 {\bf (iii)} in Fig.\ref{Fig:orbit} when the deviation $\tilde{p}_x(x)$ of momentum $p_x(x)=\sqrt{2m\mu_x}+\tilde{p}_x(x)$ along
 the orbit is much smaller than its average value $|\tilde{p}_x|\ll
 \sqrt{2m\mu_x}$. In this case we can neglect the second order
 terms in $\tilde{p}_x$ still treating it as an operator so that
    $[\varepsilon(\hat p)-\mu]\approx -i v_x\partial_x$ where the projection of the Fermi velocity on the $x$ axis is
  $v_x=p_x/m= \pm \sqrt{p_F^2- p_z^2-p_y^2}/m$. This allows to consider the regimes when the closed orbits become so small that
 semiclassical approximation does not apply. In particular it
 allows to calculate the zeroth branch which appears when the
 distance between Andreev reflection points is much smaller than
 the coherence length $\xi$. In this way one obtains a
  system of abridged equations for the wave function envelopes
  which is called the system of quasiclassical Andreev equations. The quasicalssical
  approximation inherently misses the normal reflection of
  particles from the order parameter inhomogeneities.
  As we will see below the normal reflection is crucial to describe
  the Fermi arcs and  Majorana states at the domain wall considered. However away
  from the Fermi arcs the quasicalssical approximation matches the
  exact spectrum obtained numerically.

  To obtain analytical solutions of Andreev equations
  first let us introduce the transformation of the quasiparticle wave function $\psi = (u, v)^T$
 \begin{eqnarray}\label{Eq:transformation}
  g_+= v+u \\
  g_-=v-u
  \end{eqnarray}
  so that the system of Andreev equations acquires the form
   \begin{equation}\label{Eq:BdG1}
 \begin{pmatrix}
  h_+&\varepsilon+ c_0 p_x\\
  \varepsilon- c_0p_x&h_-\\
  \end{pmatrix} \left(g_+\atop g_- \right) =0
 \end{equation}
  where $h_{\pm}= -iv_x\partial_x \pm i c_0f(x) p_y$.

  The system (\ref{Eq:BdG1}) can be transformed to the decoupled
 second-order equations
 \begin{eqnarray}\label{Eq:BdGSOp}
 \left[\partial_x^2 + U_{0+} \cosh^{-2}x\right] g_+ = E g_+ \\
 \left[\partial_x^2 + U_{0-} \cosh^{-2}x\right] g_- = E g_-
 \label{Eq:BdGSOm}
 \end{eqnarray}
 where $E=(r/\xi)^2+\alpha^2-(p_F\varepsilon/p_x\Delta_0)^2$,
 $U_{0+}=\alpha+\alpha^2$, $\alpha= rc_0 p_y/v_x$ and $U_{0-} (p_y)= U_{0+} (-p_y)$.
 The Eq.(\ref{Eq:BdGSOp}) has eigenvalues
 $$
 E_{nq}= (1+2n_q - \sqrt{1+4U_{0+}})^2/4=  (n_q-\alpha)^2
 $$
 where $n_q\geq 0$  is integer,
 which results in the quasiclassical spectral branches
 \cite{Nakahara,Wilczek}
 \begin{equation}\label{Eq:Nakahara}
 \varepsilon_{nq}=\pm \Delta_0 \frac{|p_x|}{p_F} \sqrt{2\alpha
 n_q-n_q^2+(r/\xi)^2}.
 \end{equation}
 Note that the energy branches (\ref{Eq:Nakahara}) are twofold
 degenerate with respect to the sign change of $p_x$ projection.
The zeroth spectral branch is given by the Eq.(\ref{Eq:Nakahara})
with $n_q=0$
 \begin{equation}\label{Eq:ZBrA}
 \varepsilon_{0} (p_y,p_z)= c_0 sign(p_y)|p_x|.
 \end{equation}
 The dependencies of zeroth branch  (\ref{Eq:ZBrA})
 $\varepsilon_0=\varepsilon_0(\mu_x)$ at fixed $p_y$ and
 $\varepsilon_0=\varepsilon_0(p_y)$ at fixed $p_z$ are shown in
 Fig.\ref{Fig:Emux} by red dashed lines and in Fig.\ref{Fig:Epy} by
 blue dashed lines correspondingly. At $p_y=0$ this spectral branch is discontinuous and merges
 the edge of continuum
 $\varepsilon_c=\pm c_0\sqrt{p_F^2-p_z^2}$
 at $p_y=\pm 0$.

  Substituting the
  energy of zeroth branch into the Eq.(\ref{Eq:AndreevRef})
  for the coordinates of Andreev reflection points we obtain
  $|x_{a1,2}|= p_F^{-1} (r/\xi) (p_F/|p_y|) \sim  p_F^{-1}$
  provided $|p_y|$ is not too small. For such a small distance
  between reflection points the semiclassical approximation is
  not valid and therefore it can not describe the zeroth branch.

 The spectral branches with $n_q>0$ given by Eq.(\ref{Eq:Nakahara})
 exist only in the limited range of parameters
 $\sqrt{p_F/\xi}\ll |p_x|<p_x^*$ where $p_x^*=|p_y|
 n_q/[n_q^2-(r/\xi)^2]$ provided that $n_q>r/\xi$. At $p_x=p_x^*$ the branch merges continuum.
 Thus behaviour us demonstrated in Figs.\ref{Fig:orbit}b where the
 energy branches (\ref{Eq:Nakahara}) with $n_q=0,1$ are shown by
 dashed red and blue lines as function of
 $\mu_x= \mu-(p_y^2+p_z^2)/2m$ for the parameters $r=\xi/2$ (width of the domain
 wall), $c_0 /v_F= 2.5\;10^{-2}$ and $p_y= 0.5 p_F$. The zeroth
 branch merges continuum at $p_y=\pm 0$ where the spectrum
 Eq.(\ref{Eq:ZBrA}) is discontinuous.  This behaviour is shown by dashed blue lines in
 Fig.(\ref{Fig:Epy}).

  \subsubsection{Exact spectrum near the Fermi arcs.}
  \label{Subsubsec:ExactSp}
\begin{figure}[htb]
 \centerline{\includegraphics[width=1.0\linewidth]{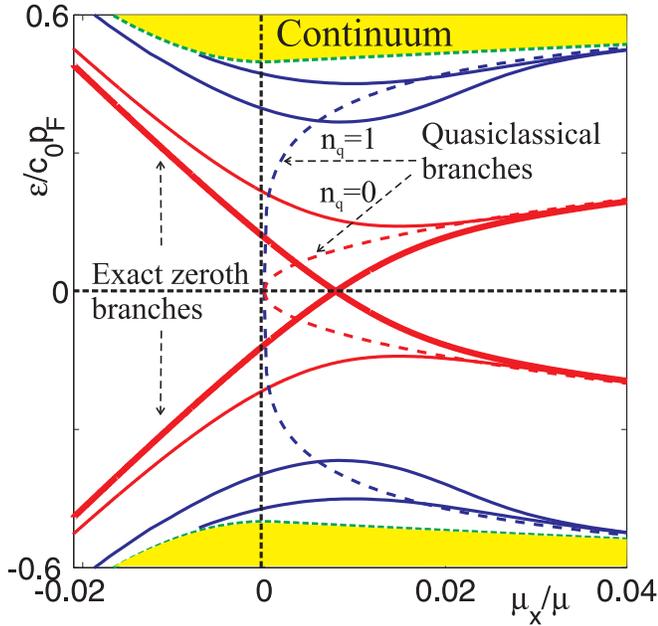}}
 \caption{\label{Fig:Emux} (Color online)
 The spectrum of surface states at the A phase
 domain wall and the AB
 interface $\varepsilon=\varepsilon(\mu_x)$
 where $\mu_x=\mu-(p_y^2+p_z^2)/2m$.
  The parameters are $r=\xi/2$ (width of the domain
 wall) and $c_0/v_F= 2.5\;10^{-2}$, $|p_y|=0.5 p_F$.
 Shown by solid red and blue lines are the spectral branches
 determined by the exact solution of BdG Eq.(\ref{Eq:Hamiltonian}).
 Thick red solid lines show the zeroth branches at $p_y=\pm 0.5 p_F$.
 Dashed lines represent quasiclassical energy branches
 (\ref{Eq:Nakahara}) for $n_q=0$ (red line) and $n_q=1$ (blue line)
  degenerate by $sign (p_x)$. The degeneracy by $sign (p_x)$ is
 removed at the region $|\mu_x|\ll \mu$ and $\mu_x<0$ where each quasiclassical branch
 splits by two modes. In accordance with general topological argument the only one zeroth branch remains
 at fixed value of $p_y$ (and spin projection $\sigma_x$ for the domain wall of the A phase) shown by the red solid line.
  The spectrum of delocalized states is shown by yellow shading and the edge of continuum
  by thin dotted line.}
 \end{figure}

 As pointed out by Nakahara \cite{Nakahara} all quasiclassical branches (\ref{Eq:ZBrA}) formally contain zero
 energy states at $p_x=0$ but Eq.(\ref{Eq:ZBrA}) is applicable if $(p_x/p_F)p_x\xi\gg 1$.
 In the limit $p_x\rightarrow 0$ and arbitrary values of $p_y$ and $p_z$ the spectrum can be
 found numerically by solving the eigenvalue problem for the
 Eq. (\ref{Eq:BdG1}). The resulting several lowest energy spectrum branches $\varepsilon=\varepsilon(\mu_x)$ are shown in the
 Fig.\ref{Fig:Emux} by solid red and blue lines as the
 function of $\mu_x= \mu-(p_y^2+p_z^2)/2m$.
  Remarkably the twofold degeneracy $p_x\rightarrow -p_x$ of the
 quasiclassical spectrum is broken at $0<\mu_x\ll \mu$ and
  at $\mu_x<0$. Therefore each of the quasiclassical energy branches (\ref{Eq:ZBrA})
 found by Nakahara splits by two modes as $\mu_x\rightarrow +0$ (and $p_x=\sqrt{2m\mu_x}\rightarrow 0$).
 The splitting is caused by normal reflection of
 quasiparticles from the order parameter spatial inhomogeneity. The normal reflection is particularly
 important for the quasiparticle trajectories passing almost
 parallel to the domain wall plane when $p_x\rightarrow 0$. In this
 case the normal reflection leads to the drastic change in the spectral branches behaviour near the zero energy point
 shown in Fig.\ref{Fig:Emux} (compare the dashed and solid lines).
 The influence of normal reflection on the spectrum of bond fermions was analyzed
 above in the framework of semiclassical approach in the section
 (\ref{Subsubsec:Semiclassical}). Comparing the Figs.(\ref{Fig:orbit}b) and (\ref{Fig:Emux}) one can see that
 the semiclassical approach qualitatively describes the doubling of spectral
 branches due to the normal reflection but gives rather large quantitative
 discrepancy with the exact spectrum.

 In accordance with general topological argument in Sec. (\ref{Subsec:Topology}) the only one zeroth branch remains
 at each fixed value of momentum projection $p_y$ (and spin
 projection $\sigma_x$ for the domain wall of the A phase).
 Exact zeroth branches for $p_y=\pm 0.5p_F$ are shown in the Fig.\ref{Fig:Emux} by thick red solid lines.
  The spectrum in Fig.\ref{Fig:Emux} is invariant with respect to the transformation
 $\varepsilon (p_y,\sigma)=-\varepsilon(-p_y,-\sigma)$ where
 $\sigma=\sigma_x$ for the case of domain wall and $\sigma=\sigma_\perp$
 for that of the AB interface. Note that in contrast to the
 quasiclassical zeroth branch shown by red dashed line the exact
 zeroth branch intersects at the Fermi level at finite value of
 $\mu_{x}=\mu_{x0}>0$. This crossing point determines the positions of
 Fermi arcs  shown in the Fig.(\ref{Fig:BoundaryStates}b) by blue and green solid lines.
 For example the Fermi arcs cross the line $p_z=0$ at $p_y=\pm p_{y0}$ where
 $p_{y0}= \sqrt{2m(\mu-\mu_{x0})}<p_F$.

 Near the projection of Weyl point on $p_yp_z$ plane in the limit $p_y\rightarrow 0$ and $|p_z|\rightarrow p_F$
  the behaviour of zeroth branch can be found analytically using
  the approach of Ref.(\onlinecite{Volovik1992}). We will treat the first term in the Hamiltonian (\ref{Eq:Hamiltonian})
 as perturbation. The rest of the terms form zero order Hamiltonian
 \begin{equation}\label{Eq:Hamiltonian0}
\hat H_0=\hat\tau_1 \hat p_x - \hat\tau_2 p_y f(x)
\end{equation}
where $\hat p_x=-i\partial_x$. We assume the model form of the
domain wall $f(x)=\tanh(x/r)$. The Hamiltonian
(\ref{Eq:Hamiltonian0}) has zero energy eigen state with the wave
function components $\psi_0=(u_0,v_0)$
 \begin{eqnarray}\label{}
 u_0=0 \\
 v_0=N^{-1/2} \cosh^{-\alpha}(x/r)
 \end{eqnarray}
 where
  $\alpha=r p_y$ and $N=\int_{-\infty}^{\infty} \cosh^{-2\alpha}(x/r) dx$.
  The perturbation of the zero energy level is given by
\begin{equation}\label{Eq:SpectrumQM}
\varepsilon_{0}(p_x,p_y)=\frac{\bar{p}_x^2}{2m}-\mu_x
\end{equation}
  where $\bar{p}_x^2 = \langle \psi_0 |\hat
 p_x^2|\psi_0\rangle$. The Fermi arc is then given by
 \begin{equation}\label{Eq:FermiArcQM}
 p_z^2=p_F^2-\bar{p}_x^2- p_y^2
 \end{equation}
 In the limit $p_y\rightarrow 0$ we obtain
 $\bar{p}_x^2=p_y^2$ so that the Fermi arc is given by
 $ p_z^2=p_F^2-2 p_y^2$ and ends at $p_z=\pm p_F$ in accordance with general topological arguments.
 The Eq.(\ref{Eq:SpectrumQM}) describes analytically the zeroth
  branch of the spectrum shown by thick red lines in
  Fig.(\ref{Fig:Emux}) near the projections of Weyl points on $p_yp_z$ plane.

\begin{figure}[htb]
 \centerline{\includegraphics[width=0.95\linewidth]{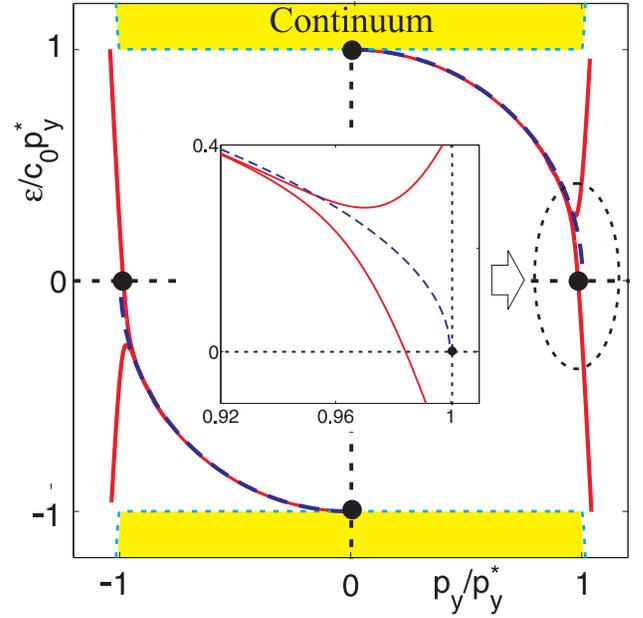}}
 \caption{\label{Fig:Epy} (Color online)
 The lowest energy spectral branches of the surface states at
 the domain wall of the A phase and at the AB
 interface as functions of $p_y$
 at fixed $p_z$. Shown by blue dashed line is
 the quasiclassical zeroth energy branch (\ref{Eq:ZBrA}) which ends at $p_y=\pm p_y^*$ where
 $p_y^*=\sqrt{p_F^2-p_z^2}$. Exact branches shown by solid red lines continue at $|p_y|>p_y^*$.
 At $p_y=\pm 0$ the discrete spectrum merges with continuum shown by the yellow shading.
 Inset: zoomed area demonstrating in detail the
 splitting of quasiclassical zeroth branches and the behavior of exact branches near the end point of
 quasiclassical one. }
 \end{figure}

  The splitting of quasiclassical zeroth branches
   $\varepsilon=\varepsilon_0(p_y) $ at fixed value of $p_z=0$ is shown in
  Fig.\ref{Fig:Epy}. Red solid lines is the exact
  spectrum obtained numerically and blue dashed lines are the
  quasiclassical zeroth branches (\ref{Eq:ZBrA}). Away from the level $\varepsilon=0$
   the correspondence of quasiclassical and exact
  spectra is of a very good accuracy. The splitting takes place in
 the  small vicinity of $p_y=\pm p^*_y$ where the quasiclassical
  branch intersect the Fermi level. The behaviour of spectral
  branches near this point is shown in the zoom inset in the
  Fig.(\ref{Fig:Epy}). All branches are discontinuous at $p_y=0$
  where they merge the continuum. Shown in Fig.(\ref{Fig:Epy}) the
  spectrum of bound fermionic states at the domain wall in $^3$He-A at $p_z=0$ coincides with the spectrum of
  electronic states at domain walls in $p+ip$ superconductor $Sr_2RuO_4$.

\section{Discussion}
\label{Sec:Discussion}

In contrast to the Fermi arcs on $^3$He-A
surface\cite{Tsutsumi2011} the Fermi arcs on the domain wall in
$^3$He-A and on the AB interface can not be obtained in the
framework of quasiclassical approximation. Indeed the
quasiclassical approximation when applied to these systems yields
that any of the subgap spectral branches given by
Eq.(\ref{Eq:Nakahara})
 intersects the Fermi level\cite{Nakahara}. Thus the number of zero energy modes in
quasiclassics is model depended and does not satisfy the index
theorem. The reason behind this discrepancy is that the zero modes
exist for $p_x=0$ (see dashed lines in Fig.\ref{Fig:orbit}b) which
corresponds to the quasiclassical trajectories almost parallel to
the plane of domain wall when one needs to get into account the
normal reflection of qasiparticles from the order parameter
inhomogeneity along the $x$ axis. Since the normal reflection is
inherently missing in quasiclassical Andreev equations (without
diagonal potential) this approximation fails to describe zero
modes and Fermi arcs on the domain wall in $^3$He-A and the AB
interface.

The modification of quasiclassical results due to the normal
reflection can be analyzed by employing the semiclassical
approximation (see Section \ref{Subsubsec:Semiclassical}) which
yields three possible types of closed orbits in $(p_x,x)$ phase
space. In case when quasiclassical approximation is valid the
regime {\bf (ii)} of only Andreev reflection is realized with two
separate orbits at $p_x>0$ and $p_x<0$ shown by the green solid
lines in Fig. \ref{Fig:orbit}a. The Bohr-Sommerfeld quantization
(\ref{Eq:BS}) yields the same energy of bound states for the two
orbits which is degenerate by $sign(p_x)$ and coincides well with
quasiclassical result for $n_q>0$. However for smaller $\mu_x$ two
Andreev orbits merge into a single one shown in
Fig.\ref{Fig:orbit}a by blue dash-dotted line. This orbit has both
Andreev and normal reflection points and therefore is not
described by the quasiclassical approximation. Moreover merging of
Andreev orbits doubles the orbit area. Therefore the number of
subgap states according to Bohr-Sommerfeld quantization is doubled
which is manifested by the appearance of the second semiclassical
branch in Fig.\ref{Fig:orbit}b simultaneously with merging of two
Andreev orbits in Fig.\ref{Fig:orbit}a. In exact spectrum obtained
by solving numerically the BdG system the branch doubling is
manifested in splitting of quasiclassical branches in
Figs.\ref{Fig:Emux},\ref{Fig:Epy}. Thus the branch splitting is
the direct result of normal reflection of quasiparticles moving
almost parallel to the domain wall.

Affected by the enhanced normal reflection the behaviour of the
energy branches changes in $p_x\rightarrow 0$ limit qualitatively
compared to the quasiclassical result. At fixed value of momentum
projection $p_y$ (and spin projection $\sigma_x$ in case of A
phase domain wall) there is only one branch left which intersects
the Fermi level and supports Majorana states in accordance with
the topological index theorem. The rest of spectral branches
become non-monotonic at small values of $|p_x|=\sqrt{2m\mu_x}$ at
$\mu_x>0$ and turn upwards at $\mu_x<0$ as can be seen in the
Fig.\ref{Fig:Emux}.

\section{Conclusion}
\label{Sec:Conclusion}

We demonstrated that Fermi arcs on the domain wall in the A-phase
and at the interface between the A-phase and the B-phase of
superfluid $^3$He obey the index theorem, which connects
topological properties of bulk states on two sides of the walls
with topology of zero energy bound states. In contrast to the
other systems supporting Fermi arcs the bulk-edge correspondence
here can be established only beyond the quasiclassical
approximation by solving the exact BdG system numerically. The
domain wall in A-phase contains 4 Fermi arcs, if one takes into
account spin degrees of freedom, while the AB interface contains 2
Fermi arcs, i.e. the same number as the surface of the A-phase.
The reason for this difference is that the Fermi arc is determined
by the Weyl points in the bulk states on two sides of the wall.
The $^3$He-B, though being a topological superfluid, does not
 contain  Weyl points and thus does not contribute to the number of the Fermi arcs.
 Note that the spectrum of localized fermions at the domain wall
 of the A phase is identical to the spectrum of electronic
 states bound at the domain wall in $p+ip$ superconductor
 $Sr_2RuO_4$.

 Majorana fermions living on the Fermi arc at the AB interface
may give an additional contribution to the calculated friction
force  acting on the moving interface \cite{Kopnin1987} at very
low temperatures. This is important for the development of the
Kelvin-Helmholtz instability of the moving AB interface observed
in Ref. \onlinecite{Blaauwgeers2002} (see also Chapter 27  of Ref.
\onlinecite{Volovik2003}).
 The latter instability has an analogue with the ergoregion instability of black holes discussed in Chapter 32 of Ref.
\onlinecite{Volovik2003}.

We considered the simplest most symmetric realizations of the A-A and A-B interfaces. The problem for future
investigations is weather the Fermi arc survives or not, if the symmetry of the interface is violated.
For that the relative homotopy group formalism applied to Green's function \cite{VayrynenVolovik2011}
is probably required.

\section{Acknowledgements}
 This work was supported, in part  by the Academy of Finland (Centers
of Excellence Programme 2012-2017), the
EU 7th Framework Programme (FP7/2007-2013, Grant No.
228464 Microkelvin), by Russian Foundation for Basic Research
 Grant No. 11-02-00891-a,  Presidential RSS Council (Grant No. MK-4211.2011.2), by Programs of RAS
 ``Quantum Physics of Condensed Matter'' and
 ``Strongly correlated electrons in semiconductors, metals, superconductors and magnetic materials''.

\end{document}